\documentclass{article}

\usepackage{PRIMEarxiv}

\usepackage[utf8]{inputenc}
\usepackage[T1]{fontenc}
\usepackage{lmodern}
\usepackage{float}

\usepackage[table,dvipsnames]{xcolor}
\usepackage{amsmath,amsfonts}
\allowdisplaybreaks 
\usepackage[section]{placeins}
\usepackage{algorithmicx}
\usepackage[ruled,vlined]{algorithm2e}

\usepackage{graphicx}
\usepackage{mathtools}
\usepackage{mathrsfs}
\usepackage{bm}
\usepackage{caption}
\usepackage{amsthm}
\usepackage{amssymb}
\usepackage[colorlinks=true]{hyperref}
\hypersetup{
  linkcolor=black,
  citecolor=black,
  urlcolor=[RGB]{5,115,185}
}
\usepackage{microtype}
\usepackage{multirow}						
\usepackage{tabularx} 						
\usepackage{booktabs}
\usepackage{array}
\usepackage{pifont}
\usepackage{colortbl}
\usepackage{stmaryrd}
\usepackage{nicefrac}
\usepackage{authblk}
\usepackage{fancyhdr}
\usepackage{enumitem}
\usepackage{footmisc}
\usepackage{sidecap}
\usepackage{titlesec}

\definecolor{LightGray}{rgb}{0.8,0.8,0.8}
\definecolor{LightCyan}{rgb}{0.74,0.83,0.9}
\definecolor{DarkBlue}{rgb}{0,0.28,0.67}
\definecolor{blizzardblue}{rgb}{0.67, 0.9, 0.93}
\definecolor{inchworm}{rgb}{0.7, 0.93, 0.36}
\definecolor{coralred}{rgb}{1.0, 0.25, 0.25}
\definecolor{celadon}{rgb}{0.67, 0.88, 0.69}

\DisableLigatures{encoding = *, family = * }

\newcommand{\sherpa}{\href{https://www.sherpa.ai/}{\textcolor{DarkBlue}{Sherpa.ai} }}

\newcommand{\cmark}{\textcolor{green!60!black}{\ding{51}}}   
 
\newcommand{\xmark}{\textcolor{red}{\ding{55}}}

\newcolumntype{Y}{>{\centering\arraybackslash}X}
\newcolumntype{Z}{>{\raggedright\arraybackslash}p{3.5cm}}
\newcolumntype{L}{>{\raggedright\arraybackslash}X}
\newcolumntype{A}[1]{>{\hspace*{-#1}\centering\arraybackslash}X}
\newcolumntype{B}{>{\centering\arraybackslash}p{5cm}}
\renewcommand{\arraystretch}{1.3}  

\definecolor{DarkColor}{gray}{0.75}			
\definecolor{LightColor}{gray}{0.9}
\definecolor{LightGrey}{rgb}{0.98,0.98,0.98}
\definecolor{DarkGrey}{rgb}{0.83,0.83,0.83}
\definecolor{BaseColor}{rgb}{0.10,0.10,0.20}
\definecolor{TextColor}{RGB}{58,88,119}
\definecolor{LightTextColor}{RGB}{229,233,205}
\definecolor{DarkTextColor}{RGB}{25,29,1}
\definecolor{NeutralBg}{rgb}{0.92,0.92,0.92}
\definecolor{LightYellow}{RGB}{255,255,102}
\definecolor{DarkOrange}{RGB}{255,90,0}
\definecolor{Green}{RGB}{0,128,0}
\definecolor{White}{gray}{1}

\newtheorem{remark}{Remark}

\usepackage{etoolbox}
\makeatletter
\patchcmd{\@begintheorem}{\textit}{\textbf}{}{}
\makeatother

\numberwithin{equation}{section}
\numberwithin{theorem}{section}
\numberwithin{proposition}{section}
\numberwithin{remark}{section}
\numberwithin{problem}{section}
\numberwithin{subsection}{section}

\usepackage{etoolbox}

\AtBeginEnvironment{tabular}{\small}

\begin{document}
\raggedbottom
\fancypagestyle{firstpagestyle}{%
  \fancyhf{}%
  \fancyhead[R]{\includegraphics[scale=1.0]{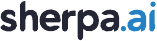}} 
  \fancyfoot[C]{\thepage}%
  \renewcommand{\headrulewidth}{0pt}%
  \renewcommand{\footrulewidth}{0pt}%
}





\title{Sherpa.ai Privacy-Preserving Multi-Party Entity Alignment without Intersection Disclosure for Noisy Identifiers}


\author{{\LARGE \href{https://sherpa.ai/}{Sherpa.ai}}}
\affil[]{research@sherpa.ai}

\maketitle
\thispagestyle{firstpagestyle}

\begin{abstract}

Federated Learning (FL) enables collaborative model training among multiple parties without centralizing raw data. There are two main paradigms in FL: Horizontal FL (HFL), where all participants share the same feature space but hold different samples, and Vertical FL (VFL), where parties possess complementary features for the same set of samples. A prerequisite for VFL training is privacy-preserving entity alignment (PPEA), which establishes a common index of samples across parties (alignment) without revealing which samples are shared between them. Conventional private set intersection (PSI) achieves alignment but leaks intersection membership, exposing sensitive relationships between datasets. The standard private set union (PSU) mitigates this risk by aligning on the union of identifiers rather than the intersection. However, existing approaches are often limited to two parties or lack support for typo-tolerant matching.

In this paper, we introduce the \sherpa multi-party PSU protocol for VFL, a PPEA method that hides intersection membership and enables both exact and noisy matching. The protocol generalizes two-party approaches to multiple parties with low communication overhead and offers two variants: an order-preserving version for exact alignment and an unordered version tolerant to typographical and formatting discrepancies. We prove correctness and privacy, analyze communication and computational (exponentiation) complexity, and formalize a universal index mapping from local records to a shared index space. This multi-party PSU offers a scalable, mathematically grounded protocol for PPEA in real-world VFL deployments, such as multi-institutional healthcare disease detection, collaborative risk modeling between banks and insurers, and cross-domain fraud detection between telecommunications and financial institutions, while preserving intersection privacy.

\end{abstract}


\begin{figure}[htp]
	\begin{center}
		\includegraphics[width=0.9\columnwidth]{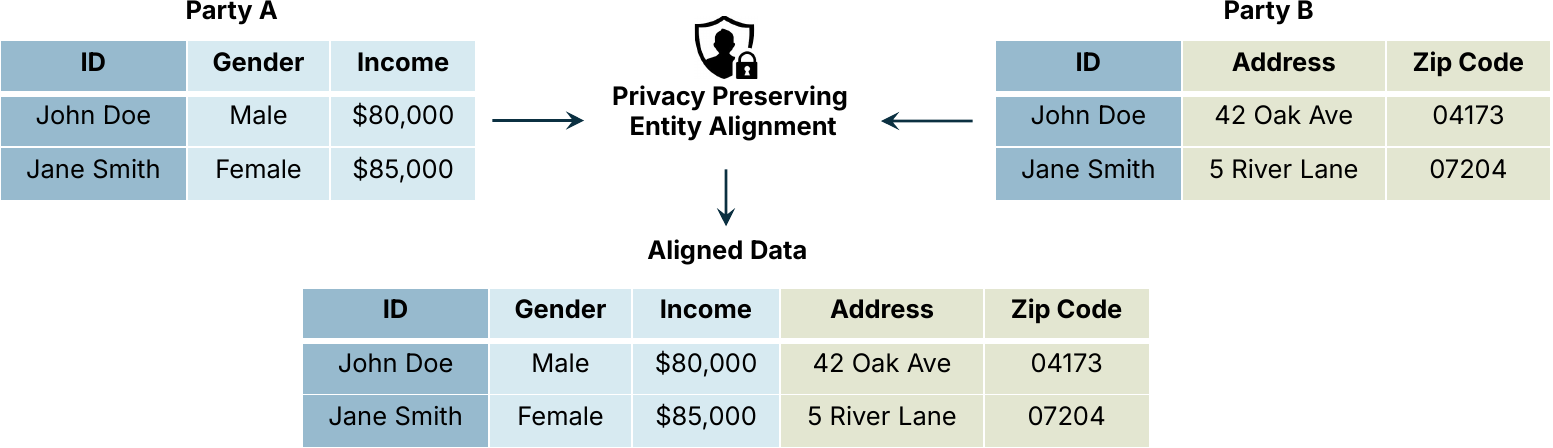}\\
		\caption{Illustrative example of entity alignment in VFL: based on the ID, Parties A and B perform private matching and produce Aligned Data without disclosing intersection membership or sensitive values.}\label{entity_alignment}
	\end{center}
\end{figure}


\section{Introduction}

Federated Learning (FL)~\cite{mcmahan2017communication} enables multiple parties (nodes, clients, or devices) to jointly train a model without sharing their raw data; instead, they exchange model parameters or updates, thereby avoiding the need to centralize datasets as in standard Machine Learning (ML). FL is commonly divided into \emph{Horizontal FL (HFL)}~\cite{yang2022horizontal}, where participants have the same features but different records, and \emph{Vertical FL (VFL)}, where participants own complementary feature sets for an overlapping population of samples~\cite{wen2023survey}.

A fundamental prerequisite in VFL is that datasets be aligned \emph{row-wise}~\cite{zhao2024deep}: (i) each record across parties refers to the same real-world entity, and (ii) records appear in the same order for all parties. In practice,
this means establishing a common index of entities while avoiding disclosure of which entities are shared across parties (i.e., intersection membership)~\cite{wang2024efficient}. For example, consider a collaboration between a bank and an insurance company: the bank holds financial transaction data, while the insurer maintains policy and claim records. To jointly train a predictive model, such as estimating default risk or detecting fraud, their datasets must be aligned so that each row corresponds to the same customer across both organizations. Achieving this alignment without revealing which customers are shared between them is the challenge that privacy-preserving entity alignment (PPEA) aims to address (see Figure~\ref{entity_alignment}).

\begin{figure}[htp]
	\begin{center}
		\includegraphics[width=1.0\columnwidth]{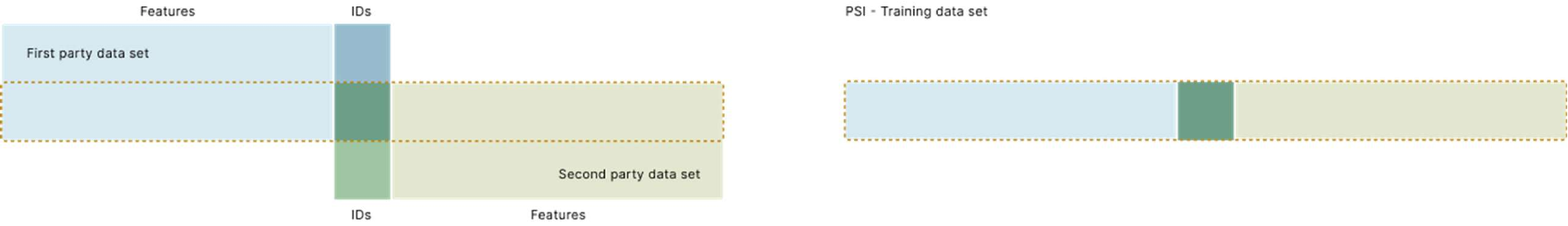}\\
		\caption{Illustration of the PSI protocol. Only the common identifiers (IDs) between the two parties are used to form the shared training dataset.}\label{fig1}
	\end{center}
\end{figure}

\begin{figure}[htp]
	\begin{center}
		\includegraphics[width=1.0\columnwidth]{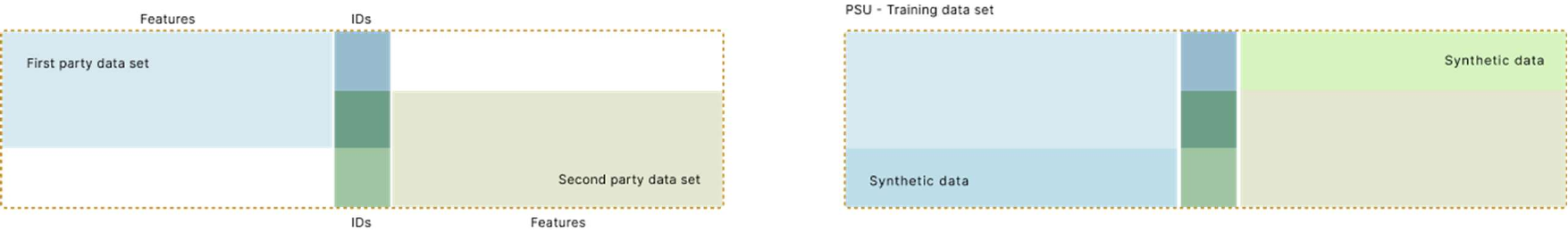}\\
		\caption{Illustration of the PSU protocol. All unique IDs across parties form the union dataset used for training; missing features for non-overlapping records are completed with synthetic data.}\label{fig2}
	\end{center}
\end{figure}

PPEA~\cite{gkoulalas2021modern} addresses this need by aligning datasets held by different parties while preserving privacy. Two main cryptographic approaches are commonly used: private set intersection (PSI) and private set union (PSU), as illustrated in Figures~\ref{fig1} and~\ref{fig2}. 

In this paper, we introduce \sherpa PSU for VFL, a multi-party, union-based entity-alignment method that employs a commutative-encryption PSU protocol to conceal intersection membership and construct a shared universal index across parties. Unlike prior PSU approaches that are primarily limited to the two-party setting and exact identifier matching, our method generalizes naturally to multiple parties and supports both exact and noisy matching regimes. After alignment on the union, missing attributes can be completed using synthetic data as a common practice used in PSU for VFL training. 

In VFL, we distinguish two conceptually distinct stages. First, at the alignment stage, PSU conceals intersection membership while constructing a private universal index across parties. Second, at the training stage, any downstream VFL method can operate on the aligned data produced by PSU. For example, paradigms such as Sherpa.ai Blind Vertical Federated Learning (SBVFL)~\cite{acero2025sherpa}, which replace true labels with server-generated synthetic labels and reduce party--server exchanges, may be employed to mitigate label and gradient leakage as well as communication overhead. In this sense, PSU is independent of the subsequent training method and serves as a privacy-preserving preprocessing step for VFL. 

\subsection{Motivation and Challenges}

PSI privately computes the intersection of the parties’ identifier sets (see Figure~\ref{fig1}) enabling them to identify which identifiers they share. However, PSI outputs the intersection of identifiers as well, which can pose a privacy risk. For example, consider a federation of a bank and a cancer clinic: learning that a particular customer of the bank is in the intersection, it implies that this individual is in the clinics’s records, potentially leaking sensitive health information (i.e., the bank could infer that the customer suffers from cancer, and hence, denying them a loan). Revealing intersection membership can therefore violate privacy~\cite{sun2021vertical}; this motivates union-based alignment that keeps membership hidden.

PSU computes the union of identifier sets (see Figure~\ref{fig2}) without revealing intersection information. In PSU, the parties compute the set of all identifiers that appear in at least one dataset. By merging datasets on the union rather than the intersection, the parties avoid disclosing which identifiers they have in common. When using the union, some feature values will be missing for entities present only in one party. These gaps can be filled with synthetic data generated locally by each party, which prior work has shown can preserve model utility while improving privacy by hiding membership~\cite{sun2021vertical}.

While several PSU protocols have been proposed in recent years, they have key limitations. Two-party designs, such as Sun et al.~\cite{sun2021vertical} and Tu et al.~\cite{tu2025fast}, target balanced or unbalanced two-party settings and, thus, do not directly scale to multi-party VFL. Multi-party or high-throughput variants (e.g., Gao et al.~\cite{gao2023toward}) prioritize efficiency with heavier primitives but do not support typo-tolerant matching. Finally, differentially private unions (e.g., Gopi et al.~\cite{gopi2020differentially}) produce approximate, rather than exact, unions. Motivated by these gaps, we develop a multi-party PSU for VFL that hides intersection membership and supports two alignment regimes: (i) an exact-hash, order-preserving regime, which is theoretically optimal when identifiers are clean and consistently formatted, and (ii) a fuzzy matching, unordered regime, which is substantially robust in real-world scenarios with noisy, heterogeneous identifier fields, but might give less accurate results in ideal conditions. In addition, some privacy-preserving entity-alignment approaches rely on a trusted coordinator or stronger trust assumptions, which may be undesirable in cross-organizational deployments; our protocol avoids this requirement. Further details and guarantees are provided in Sections~\ref{Comparison with related work} and~\ref{sec:Proposedsolution}.


\subsection{Contributions}

The main contributions of this work are summarized as follows:

\begin{itemize}
    \item We propose the multi-party \sherpa PSU protocol for PPEA in VFL. Unlike traditional PSI methods, our hashing-based PSU construction enables secure alignment across multiple parties without requiring a trusted third party and prevents the disclosure of intersection membership. 

    \item We introduce an $n$-gram tokenization preprocessing step that enhances robustness to formatting inconsistencies and typographical errors, supporting both order-preserving (exact) and unordered (noisy) matching depending on data quality.

    \item We formalize the commutative encryption process based on the Diffie–Hellman key exchange principle, generalizing existing two-party PSU methods~\cite{sun2021vertical} to a multi-party setting with provable privacy guarantees under the semi-honest model.

    \item We define the procedure for computing universal indices that map each party’s local records into a shared index space, enabling subsequent data integration and joint model training.

    
\end{itemize}

In summary, our approach establishes a mathematically grounded and robust to noisy identifiers framework for PPEA in VFL, maintaining confidentiality while remaining practical for deployment in real-world, multi-institutional collaborations.


The remainder of this paper is organized as follows. Section~\ref{Comparison with related work} reviews related work. In Section~\ref{sec:Problemformulation}, we formalize the problem setting for PSU in VFL. Section~\ref{sec:Proposedsolution} details the proposed solution, including identifier preprocessing, hashing, PSU protocols (with or without order preservation), and synthetic data generation. Finally, Section~\ref{sec:Conclusions} concludes the paper.


\section{Related Work}
\label{Comparison with related work}

Several PSU protocols have been proposed in the literature, leveraging diverse cryptographic techniques. For example, Sun et al.~\cite{sun2021vertical} proposed one of the first PSU solutions specifically for VFL without revealing intersection membership. Their protocol (which we build upon) is limited to two parties; in contrast, our approach generalizes to $P$ parties and introduces $n$-gram tokenization for improved matching. Gao et al.~\cite{gao2025pulse} developed PULSE, a parallel multi-party PSU protocol that leverages fast cryptographic operations (e.g., symmetric-key primitives and oblivious transfers) to efficiently handle large-scale datasets. Our approach shares a similar multi-party setting but prioritizes minimal communication rounds over parallel throughput. Tu et al.~\cite{tu2025fast} present an enhanced two-party PSU protocol that supports both balanced and unbalanced set sizes, achieving better computational and communication performance than earlier methods. In contrast, our protocol minimizes communication overhead by utilizing commutative encryption and naturally accommodates multiple parties without requiring additional assumptions.

Beyond purely cryptographic methods, some works rely on differential privacy (DP). For instance, Gopi et al.~\cite{gopi2020differentially} design algorithms that produce the union with rigorous privacy guarantees by injecting noise. Such DP-based approaches do not reveal exact intersection membership, but they trade off some accuracy (and typically do not yield \emph{exact} unions) in exchange for strong privacy. Other approaches employ \emph{homomorphic encryption (HE)}: for example, Tu et al.~\cite{tu2023fast} use fully HE (FHE) to compute an unbalanced PSU, achieving strong security with relatively heavy computation. Compared to these, our protocol avoids expensive public-key operations on large data, instead using hashing and modular exponentiations that are efficient, and requires only a few rounds of communication. Our primary design goal is to minimize both the number of communication rounds and the amount of exchanged data. The trade-off is that the ``PSU without order'' variant of our method involves a potentially expensive comparison step for handling typos, which can be mitigated through low-level implementations or optimized data structures such as Bloom filters. For additional references on the topic, we refer the reader to \cite{kissner2005privacy,frikken2007privacy,seo2012constant,kolesnikov2019scalable,jia2022shuffle,zhang2023linear,gao2024multi,gao2023toward,jia2024scalable,zhang2024unbalanced}.

Our approach builds upon and extends prior PSU research. In particular, our multi-party commutative encryption scheme generalizes the two-party PSU method of Sun et al.~\cite{sun2021vertical} by supporting any number of parties and introducing $n$-gram-based noisy matching. Unlike many PSI/PSU protocols that rely on heavier cryptographic tools (e.g., oblivious transfer, garbled circuits, or FHE)~\cite{jimenez2025security}, our method keeps the computations relatively lightweight (modular exponentiations and hashing) and aims to reduce communication to two main rounds. Recent works, such as Gao et al.~\cite{gao2025pulse} and Tu et al.~\cite{tu2025fast}, focus on optimizing PSU for performance, achieving notable speed-ups through parallel operations and specialized data structures, albeit at the cost of increased protocol complexity. In contrast, our protocol emphasizes ease of integration within an FL system and simplicity of implementation, assuming semi-honest parties.

Beyond PSU-specific research, complementary lines of work are directly relevant to PPEA in VFL. Private Sample Alignment (PSA) protocols have been explored to achieve reliable multi-client VFL deployments and scalable two-party settings~\cite{xi2025private,wang2025psa}, offering alternative building blocks to PSU. For asymmetric federations, differential PSI (DPSI) protects membership by adding calibrated noise to the revealed results~\cite{he2022differentially}. Noisy or approximate matching has been studied in PSI for biometric search~\cite{uzun2021fuzzy}, while classic privacy-preserving record linkage (PPRL) methods based on $n$-grams and Bloom filters~\cite{schnell2009privacy,durham2013composite} are adapted to our unordered (noisy) matching design.

A concise comparison of privacy properties across standard entity alignment (PSI), standard PSU, and our proposed \sherpa PSU is presented in Table~\ref{tab:entity_resolution_privacy}. Standard entity-alignment (PSI) protocols typically reveal the intersection between parties and are limited to exact, pairwise alignment, whereas standard PSU constructions hide the intersection but do not support multi-party execution and assume perfectly matching identifiers. In contrast, our proposed \sherpa PSU simultaneously (i) computes only the union without revealing the intersection, (ii) naturally extends to more than two parties via a commutative-encryption design, and (iii) supports privacy-preserving noisy alignment to cope with typos and non-canonical identifiers. This combination of properties, which is crucial in realistic VFL deployments, is not provided by existing PSI/PSU-based entity-alignment schemes.

\begin{table}[H]
  \scriptsize
  \setlength{\tabcolsep}{4pt}
  \renewcommand{\arraystretch}{1.2}
  \centering
  \begin{tabularx}{\textwidth}{L|YYY}
    \toprule
    & \begin{tabular}[c]{@{}c@{}}\textbf{Standard Entity} \\ \textbf{Alignment (PSI)}\end{tabular} & \begin{tabular}[c]{@{}c@{}}\textbf{Standard} \\ \textbf{PSU}\end{tabular} & \begin{tabular}[c]{@{}c@{}}\textbf{\sherpa} \\ \textbf{PSU}\end{tabular} \\ 
    \midrule
    Raw identifiers exchanged & \cmark\ No & \cmark\ No & \cmark\ No \\ 
    Intersection revealed     & \xmark\ Yes & \cmark\ No & \cmark\ No \\
    Multi-party     & \cmark\ Yes & \xmark\ No & \cmark\ Yes \\    
    
    Noisy alignment     & \xmark\ No & \xmark\ No & \cmark\ Yes \\    
    \bottomrule    
  \end{tabularx}
  \caption{Comparison of the privacy properties of different entity alignment protocols.}
  \label{tab:entity_resolution_privacy}
\end{table}

\section{Problem Formulation}
\label{sec:Problemformulation}

We consider a set of $P$ parties wishing to perform VFL. Without loss of generality, assume there are $P-1$ \textit{passive parties}
\begin{equation*}
\mathscr{F}_0,\dots,\mathscr{F}_{P-2}
\end{equation*}
and an \textit{active party}
\begin{equation*}
\mathscr{F}_{P-1}.
\end{equation*}
For each $k \in \{0, \dots, P-1\}$, party $\mathscr{F}_k$ owns a dataset $\mathcal{D}_{k}=\left\{\vec{x}_k^{\,i}\right\}_{i\in \left\{0,\dots,N_k-1\right\}}$ consisting of $N_k$ data examples (rows), where each example (or row) $\vec{x}_k^{\,i}\in \mathbb{R}^{d_{k}}$, $N_k\in \mathbb{N}\setminus \left\{0\right\}$ and $d_{k}\in \mathbb{N}\setminus \left\{0\right\}$.

All parties agree on a common set of one or more identifying features that will be used for record matching (for example, a combination of name, phone number, and address). We denote by $d_{\mbox{\tiny{match}}}$ the number of such identifying features (this $d_{\mbox{\tiny{match}}}$ is the same for every party). Let
\begin{equation*}
\pi_k:\mathbb{R}^{d_{k}}\longrightarrow \mathbb{R}^{d_{\mbox{\tiny{match}}}},
\end{equation*}
be the projection that extracts the $d_{\mbox{\tiny{match}}}$ identifying attributes from party $\mathscr{F}_k$’s feature space. Each party can then derive from its dataset a set of \textbf{raw identifiers} (one identifier per example) as
\begin{equation*}
I_{\mbox{\tiny{raw}},k}\coloneqq \left\{\pi_k\left(\vec{x}_k^{\,i}\right) \ | \ k=0,\dots,N_k-1\right\}.
\end{equation*}
for $k = 0, \dots, P-1$. In other words, $I_{\mbox{\tiny{raw}},k}$ is the set of identifier tuples (such as personal data fields) for the records held by party $\mathscr{F}_k$. Our goal is to perform the following steps in a privacy-preserving manner, \emph{without} revealing to any party which identifiers are shared or not shared with other parties.

\begin{enumerate}
	\item \textbf{Compute the Union of Identifiers.} Determine the union of all parties’ identifier sets:
	\begin{equation}
	\mathcal{U} \coloneqq \bigcup_{k=0}^{P-1} I_{\mbox{\tiny{raw}},k}.
	\end{equation}
	Let $N = |\mathcal{U}|$ be the total number of unique identifiers across all parties.
	
	\item \textbf{Define Universal Indices.} Assign each identifier in the union $\mathcal{U}$ a unique \textbf{universal index}. Let $V \coloneqq \left\{0,\dots,N-1\right\}$ be the set of universal indices. This is essentially an indexing of the union $\mathcal{U}$.
	
	\item \textbf{Align Local to Universal Index Mappings.} For each party $\mathscr{F}_k$, determine a mapping
	\begin{equation}
	\varphi_k:\left\{0,\dots,N_k-1\right\}\longrightarrow V,
	\end{equation}
	which maps each \textbf{local index} $i$ (the position of a record in party $\mathscr{F}_k$’s dataset) to the corresponding universal index in $V$. In other words, if $\pi_k(\mathbf{x}_{k}^{\,i}) \in \mathcal{U}$ ends up as the $j$-th element of the union, then $\phi_k(i) = j$. This mapping $\phi_k$ allows each party to translate its local records into the unified indexing of the union.
\end{enumerate}

All of the above should be achieved without any party learning which of its identifiers were or were not present in other parties’ datasets (thus, intersection membership remains private). Next, we describe our solution to this problem.

\section{Our Proposed PSU Solution}
\label{sec:Proposedsolution}

\begin{figure}[h]
	\centering
	\includegraphics[width=0.8\linewidth]{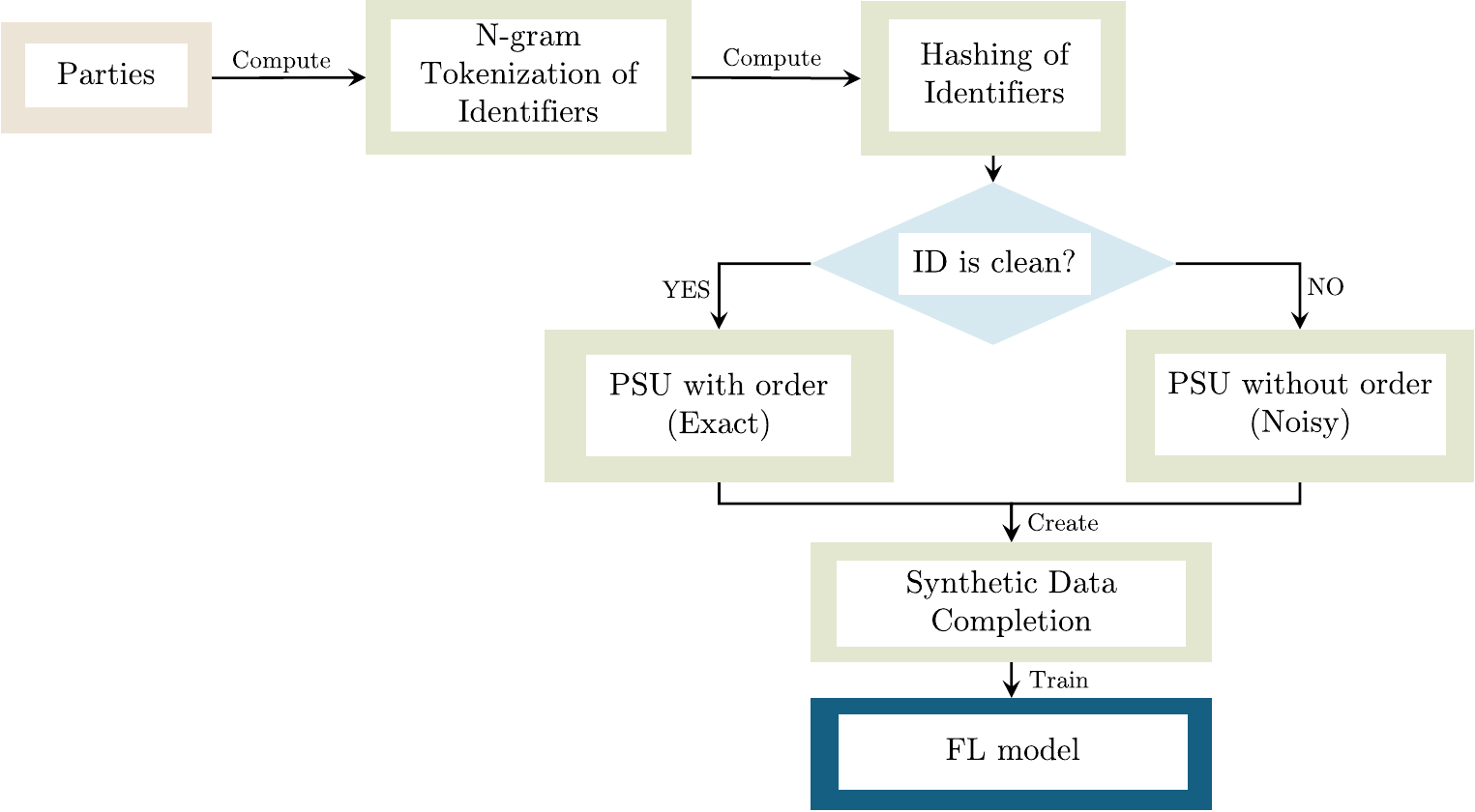}
	\caption{Pipeline of the proposed PSU protocol for multi-party VFL.}	\label{fig:psu_process}
\end{figure}

Our pipeline proceeds in four stages (see Figure~\ref{fig:psu_process}), to enable VFL PPEA without revealing intersection membership:

\begin{enumerate}
	\item \textbf{$n$-gram Tokenization of Identifiers:} We first split each identifier in $I_{\mbox{\tiny{raw}},k}$ (which may be a composite of multiple fields) into a set of overlapping substrings of length $n$, known as \emph{n}-grams. This helps standardize the format and allows for partial matching to accommodate error tolerance.
	
	\item \textbf{Hashing of Identifiers:} The $n$-grams are hashed to fixed-length values. Hashing ensures a uniform representation (e.g., fixed-size bit strings or integers) for tokens, and provides a layer of one-way protection so that plaintext identifier values are not directly used in the cryptographic protocol.
	
	\item \textbf{PSU Protocol:} Next, we perform a multi-party PSU protocol on the sets of hashed $n$-grams to obtain the union and the index mappings. Our PSU protocol uses commutative encryption in the cyclic group of quadratic residues to hide identifier values and achieve the required privacy properties.

    \item \textbf{Synthetic Data Completion:} After aligning on the union and obtaining universal indices, each party fills missing feature values for non-overlapping entities using privacy-preserving synthetic data generators, enabling downstream FL training without sharing raw data.
    
\end{enumerate}

The algorithms we propose are a generalization of \cite[Algorithm 1, page 5]{sun2021vertical}, with the following additional features.
\begin{itemize}
	\item They work for $P$ parties, where $P\geq 2$.
	\item Different identifier features are treated separately.
	\item Each identifier's feature is split in $n$-grams of assigned length.
\end{itemize}

Conceptually, we distinguish two cases for the alignment problem. First, we consider alignment with exact hashes, where each identifier is represented as an ordered sequence of $n$-grams and two records are deemed equal only if all tokens match in the same order (Section~\ref{subsec:PrivateSetUnion(PSU), with order}). This regime is theoretically optimal when all parties store clean, consistently formatted identifiers, but it is less realistic in practice, where fields may be split, reordered, or contain typos. Second, we introduce a noisy matching regime that operates on unordered multisets of $n$-grams (Section~\ref{subsec:PrivateSetUnion(PSU), without order}). This ``without order'' variant is not exact even under ideal conditions, but in practice gives a very good approximation while being substantially more robust to noisy and heterogeneous identifier fields. Both cases share the same cryptographic PSU core; they differ only in how identifiers are tokenized and compared.

In general, the PSU protocol is designed and analyzed in the group of quadratic residues modulo a safe prime. Specifically, let $p$ be a \textbf{safe prime}, i.e., $p$ is prime and $(p-1)/2$ is also prime. Denote by $\mathbb{Z}_p = \left\{[0]_p,[1]_p,\dots,[p-1]_p\right\}$ the ring of integers mod $p$, and by $\mathbb{Z}_p^* = \left\{[1]_p,[2]_p,\dots,[p-1]_p\right\}$ the multiplicative group of integers mod $p$. We define the group of quadratic residues modulo $p$ as
\begin{equation}\label{quad_residues}
QR\left(\mathbb{Z}_p^{\star}\right)\coloneqq \left\{[x]_p \ \big| \ \exists y\in \mathbb{Z}, \ y^2\equiv x \not\equiv 0  \ \left(\mbox{mod} \ p\right)\right\}\,.
\end{equation}

In other words, $QR\left(\mathbb{Z}_p^{\star}\right)$ is the subgroup of $\mathbb{Z}_p^*$ consisting of all non-zero squares mod $p$. By using the properties of the additive group $\left(\mathbb{Z}_p, +\right)$ and the cyclic group $\left(\mathbb{Z}_p^{\star}, *\right)$, it is possible to prove that $\left(QR\left(\mathbb{Z}_p^{\star}\right),*\right)$ is a group of order $\frac{p-1}{2}$. The security of the algorithms is based on the decisional Diffie-Hellman assumption \cite{diffie1976new}, which stipulates that in the discrete group \eqref{quad_residues}, performing the power (encryption) is easy, whereas performing the logarithm (decryption) is hard. As such, for our algorithms, all the operations will be taken modulo $p$ or $q=\frac{p-1}{2}$, where $p$ is a safe prime number.

\subsection{$n$-gram Tokenization of Identifiers}
\label{subsec:Separation in m-grams}

For each party $\mathscr{F}_k$ and each raw identifier $\mbox{id}\in I_{\mbox{\tiny{raw}},k}$, the identifier is decomposed into \textit{n}-grams. An identifier $\mbox{id}$ may be a single string (e.g., a customer ID) or a tuple of attributes (e.g., first name, last name, address, etc.).

Formally, for every $k=0,\dots,P-1$, party $\mathscr{F}_k$ possesses an identifiers set $I_{\mbox{\tiny{raw}},k}$. A raw identifier $\mbox{id}\in I_{\mbox{\tiny{raw}},k}$ is a vector
\begin{equation*}
\mbox{id}=\left(\mbox{id}_1,\dots,\mbox{id}_{d_{\mbox{\tiny{match}}}}\right)\in\mathbb{R}^{d_{\mbox{\tiny{match}}}}.
\end{equation*}

Without loss of generality, assume each component $\mbox{id}_r$ can be represented as a string (we can stringify numeric fields as well):
\begin{equation*}
\mathcal{S}\coloneqq \left\{\mbox{s} \ | \ \mbox{s} \ \verb!Python string!\right\}.
\end{equation*}

We can cast in-place each component of the identifier $\mbox{id}$ as a Python string. Then,
\begin{equation*}
\mbox{id}=\left(\mbox{id}_1,\dots,\mbox{id}_{d_{\mbox{\tiny{match}}}}\right)\in\mathcal{S}^{d_{\mbox{\tiny{match}}}}.
\end{equation*}

All parties $\mathscr{F}_0,\dots,\mathscr{F}_{P-1}$ agree on the length $n_r$ of the $n$-grams in which component $r$ of identifiers will be split, for each $r\in \left\{1,\dots,d_{\mbox{\tiny{match}}}\right\}$.
Moreover, in order to perform a proper $n$-grams splitting, they need to agree on a length for the string of each component of identifiers: let $L_r$ be the established length of the component $r$ of identifiers; if $n_r>L_r$, we redefine
\begin{equation*}
L_r \leftarrow n_r.
\end{equation*}

For every $k=0,\dots,P-1$, for any identifier $\mbox{id}\in I_{\mbox{\tiny{raw}},k}$, for each $r\in \left\{1,\dots,d_{\mbox{\tiny{match}}}\right\}$,
\begin{itemize}
	\item if the length of the string $\mbox{id}_r$ is smaller than $L_r$, then $L_r-\mbox{length}\left(\mbox{id}_r\right)$ empty spaces are added;
	\item if the length of the string $\mbox{id}_r$ is greater than $L_r$, then the last $\mbox{length}\left(\mbox{id}_r\right)-L_r$ characters are removed.
\end{itemize}
At this stage, we can separate $\mbox{id}_r$ in $L_r-n_r+1$ $n_r$-grams, using a sliding window approach. Namely, for $l\in \left\{0,\dots,L_r-n_r\right\}$, the $l$-th $n_r$-grams of $\mbox{id}_r$ is
\begin{equation*}
c_{r,l}\coloneqq \left(\mbox{id}_{r,l},\dots,\mbox{id}_{r,l+n_r-1}\right).
\end{equation*}

For $k=0,\dots,P-1$, set
\begin{multline*}
I_{\mbox{\tiny{separated-raw}},k}\coloneqq \bigg\{\left(c_{i,r,l}\right)_{i,r,l} \ \Big| \ i \in \left\{0,\dots,N_k-1\right\}, \ r \in \left\{1,\dots,d_{\mbox{\tiny{match}}}\right\}, \ l\in \left\{1,\dots,L_r-n_r+1\right\}\bigg\}.
\end{multline*}

The motivation for $n$-gram separation is twofold. First, it provides a flexible way to handle minor discrepancies in strings (e.g., typos or different formatting). Concretely, it allows our protocol to align identifiers such as ``123 Main St.'' stored as a single field with records where the street number and name are split across fields (``123'' and ``Main St.''), and to match ``123 Main St.'' against variants like ``123 Main Street'', ``123 main st'', or ``123 Main Str''. Similarly, it is robust to differences in capitalization (``SMITH'' vs. ``Smith''), accent marks (``José'' vs. ``Jose''), and small typographical errors. Because such representations still share most of their $n$-grams, they are treated as near-matches
in the unordered case. Second, it can improve matching accuracy by ensuring that tokens are compared at a granular level rather than via whole-string comparisons.

Our use of $n$-gram hashing is related to prior work on deep structured semantic models for web search, where $n$-gram vectors are used as inputs to neural networks, and 3-grams are found to offer a good compromise between robustness to small variations and the rate of hash collisions when order is not preserved~\cite{huang2013learning}. In our setting, such noisy matching naturally induces two types of errors: (E1) false negatives, where records that truly exist in both parties are not linked, and (E2) false positives, where records belonging to different individuals are incorrectly linked. Since E2 is typically much more harmful in privacy-preserving entity resolution, we choose the similarity threshold in our matching step to strongly penalize E2-type errors, accepting a small number of E1 errors as the cost of avoiding incorrect links. Moreover, while additional neural layers on top of $n$-gram vectors can make direct inference of the original identifiers more difficult, they are generally designed to preserve enough information to approximately reconstruct the input and should not be seen as a primary privacy mechanism. In our protocol, $n$-gram noisy matching is used to enhance robustness in record linkage, while strong privacy guarantees are provided by the subsequent encryption layer.

\subsection{Hashing of Identifiers}
\label{subsec:Identifiershashing}

After tokenization, each party hashes its $n$-grams to obfuscate their values and to enable efficient cryptographic processing. This second part of our solution addresses the hashing of $n$-grams, mapping raw feature values into the group defined in Equation \eqref{quad_residues}.

Our hashing procedure mainly consists of two steps:
\begin{enumerate}
	\item Hashing by SHA3-256 and casting to \verb!int!;
	\item Projection onto $QR\left(\mathbb{Z}_p^{\star}\right)$.
\end{enumerate}
The composition of the above two operations defines a mapping:
\begin{equation*}
h:\mathcal{S}\longrightarrow QR\left(\mathbb{Z}_p^{\star}\right),
\end{equation*}
where 
\begin{equation*}
\mathcal{S}\coloneqq \left\{\mbox{s} \ | \ \mbox{s} \ \verb!Python string!\right\}.
\end{equation*}

For every $k=0,\dots,P-1$, the hashed identifiers set will be denoted by $I_{k}$. We have
\begin{equation*}
I_{k}= \left\{h\left(c\right) \ \Big| \ c\in I_{\mbox{\tiny{separated-raw}},k}\right\}.
\end{equation*}

Hashing provides a layer of privacy (an adversary must invert the hash to guess the original token) and also standardizes the representation length of tokens. Most importantly, all parties use the \textbf{same hash function} $h$, so if two parties have an identical $n$-gram, they will end up with the same hash value. From this point onward, our protocol operates on these hashed identifiers.

\subsubsection{Commutative Encryption}

Our PSU protocol utilizes the group $QR\left(\mathbb{Z}_p^{\star}\right)$, defined earlier, as the space for commutative encryption. We choose a large safe prime $p$ (on the order of 2048 bits or larger for security) and let $q = (p-1)/2$ (which is prime). All operations on hashed identifiers will be performed modulo $p$, and some index arithmetic will be modulo $P$ (for party indices).

\begin{figure}[h]
	\centering
	\includegraphics[width=0.7\linewidth]{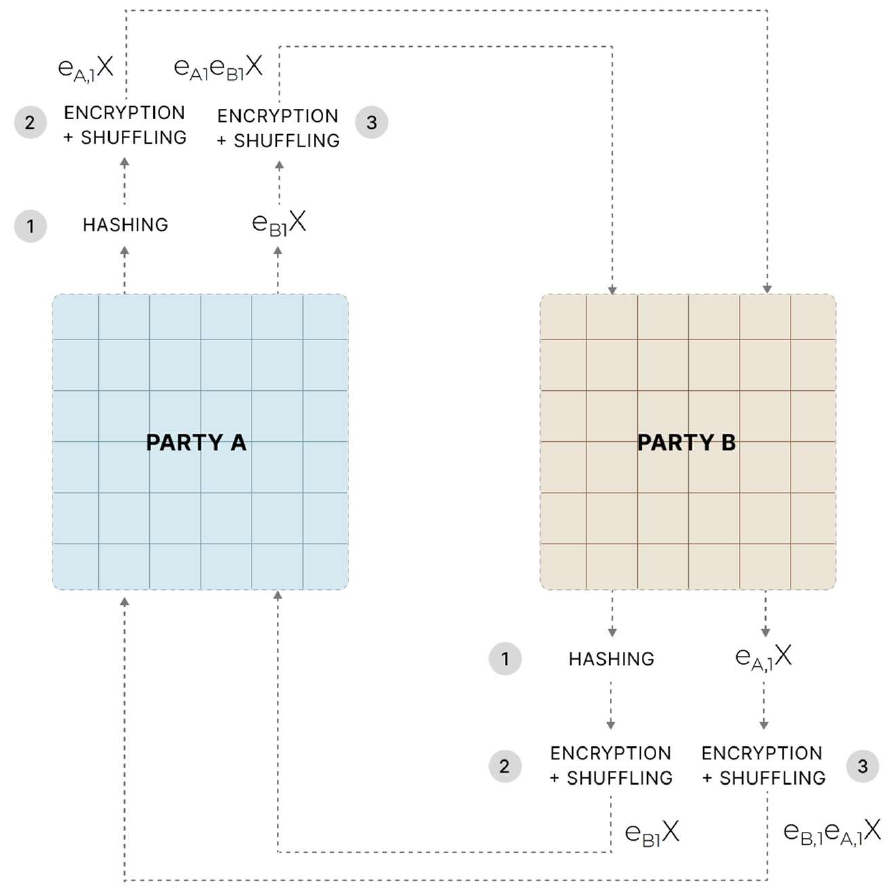}
	\caption{Scheme describing the main steps of the first part of the Diffie-Hellman protocol employed for PSU.}\label{fig:PSU_1_DF}
\end{figure}

Figure~\ref{fig:PSU_1_DF} illustrates the first phase of the commutative encryption process that underpins the PSU protocol. Each party (A and B in the example) begins by hashing its identifiers and encrypting them with its local secret exponent. The encrypted identifiers are then shuffled and exchanged between the parties. Upon receiving the counterpart’s encrypted set, each party re-encrypts the values with its own secret exponent and shuffles them again before returning them. This iterative, commutative exchange ensures that identifiers are doubly encrypted under both parties’ keys while preserving the property that identical identifiers yield identical ciphertexts, a prerequisite for performing the secure union.

Each party $\mathscr{F}_k$ generates a secret exponent $s_k$ (in practice, multiple exponents per party are used for different protocol phases, denoted $s_{k_1}, s_{k_2}, s_{k_3} \in \{0,\dots,q-1\}$). We define the encryption function $e_s$ for exponent $s$ applied on hashed token $x \in \mathbb{Z}_p$ as:

\begin{equation}
e_{s,\mbox{\tiny{one component}}}(x) = x^s \mod p\,.
\end{equation}

We extend this to an identifier’s tokens (across all features) by applying $e_s$ to each token: let an identifier be represented as a tuple $(x_{r,l})$ where $r$ is the index of the feature (from $1$ to $d_{\text{match}}$) and $l$ is the index of the $n$-gram within that feature (from $1$ to $L_r - n + 1$, assuming $L_r$ is the length of feature $r$ for that identifier). Next, we define the following vectorized encryption functions.

Let $s\in \left\{0,\dots,q-1\right\}$ and
\begin{equation*}
e_{s}:\prod_{r=1}^{d_{\mbox{\tiny{match}}}}\left[\prod_{l=1}^{L_r-n_r+1}\mathbb{Z}_p\right]\longrightarrow \prod_{r=1}^{d_{\mbox{\tiny{match}}}}\left[\prod_{l=1}^{L_r-n_r+1}\mathbb{Z}_p\right]
\end{equation*}
\begin{equation*}
\left(x_{r,l}\right)_{r\in \left\{1,\dots,d_{\mbox{\tiny{match}}}\right\}, l\in \left\{1,\dots,L_r-n_r+1\right\}}\longmapsto  \left(x_{r,l}^s\right)_{r\in \left\{1,\dots,d_{\mbox{\tiny{match}}}\right\}, l\in \left\{1,\dots,L_r-n_r+1\right\}},
\end{equation*}
where $\sigma:\left\{0,\dots,N_k-1\right\}\longrightarrow \left\{0,\dots,N_k-1\right\}$ is a random permutation.

By the commutativity of the product in $\mathbb{Z}_p$, $e_{s_1}\circ e_{s_2}=e_{s_2}\circ e_{s_1}$, for some $s_1$, $s_2$ in $\left\{0,\dots,q-1\right\}$.
Because exponentiation is applied independently to each token, two identical identifiers (same tokens in the same order) encrypted with the same exponent \(s\) yield identical outputs, while identifiers that differ in any token produce different encrypted outputs (up to hash or exponent collisions). Moreover, since \(e_{s_1} \circ e_{s_2} = e_{s_2} \circ e_{s_1}\) for any \(s_1, s_2 \in \{0,\dots,q-1\}\), applying exponents in any order is equivalent to a single exponentiation by \(s_1 s_2 \bmod q\). This commutativity is what enables our multi-party PSU construction.

From a cryptographic standpoint, this layer is not a new primitive but the standard Diffie--Hellman exponentiation in the safe-prime subgroup \(QR(\mathbb{Z}_p^\ast)\), i.e., the mapping \(x \mapsto x^s \bmod p\) for a secret exponent \(s \in \{1,\dots,q-1\}\). We deliberately use this primitive in a deterministic, commutative masking mode rather than as a randomized public-key encryption scheme. This is essential for PSU: (i) exponentiations under different parties' secret exponents must commute, so that applying all exponents in any order yields the same masked identifier, and (ii) identical identifiers must remain identical after all exponentiations, so that the server can compute the union by equality tests on the resulting masked values. Standard randomized public-key encryption schemes would typically destroy these properties, since encrypting the same value twice yields unrelated ciphertexts and encryptions under different keys do not commute. Our choice therefore, provides exactly the algebraic structure needed for multi-party PSU without a trusted third party. The privacy analysis is carried out in the semi-honest model and relies on the standard Decisional Diffie--Hellman assumption in \(QR(\mathbb{Z}_p^\ast)\). 

\begin{figure}[H]
	\centering
	\includegraphics[width=0.7\linewidth]{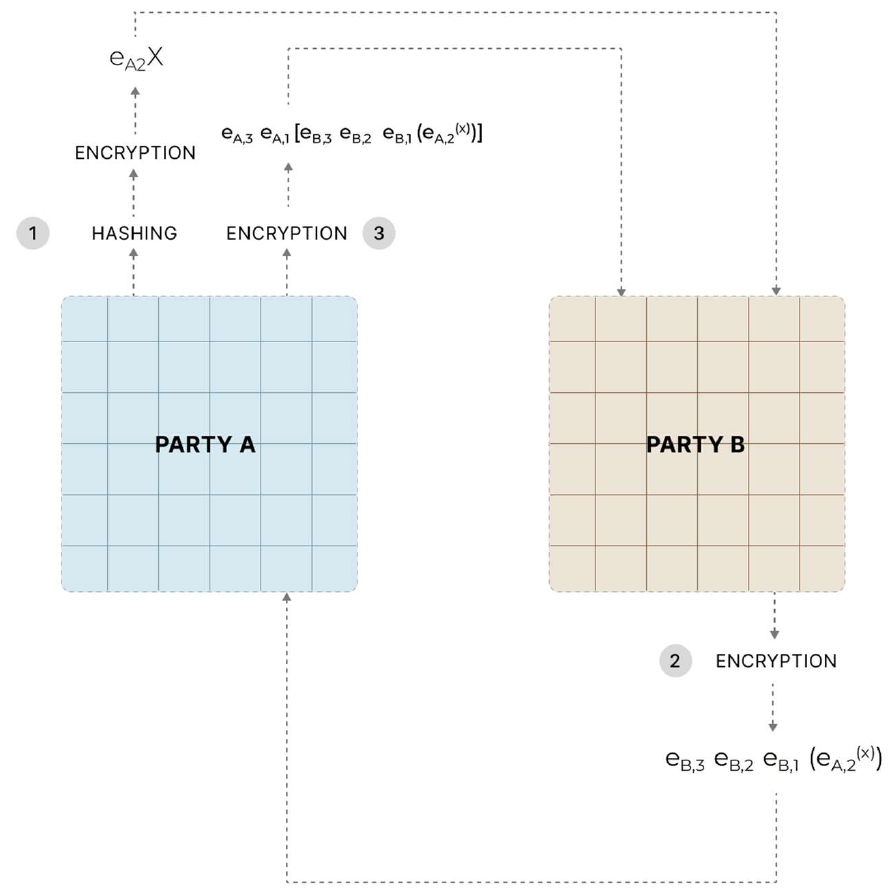}
	\caption{Scheme describing the main steps of the second part of the Diffie-Hellman protocol employed for PSU.}	\label{fig:PSU_2_DF}
\end{figure}

Figure~\ref{fig:PSU_2_DF} illustrates the second phase of the commutative encryption process, which determines the final mapping between local identifiers and their encrypted counterparts in the union. After the initial double-encryption round (Figure~\ref{fig:PSU_1_DF}), each party performs additional encryption operations using its remaining secret keys and returns the results to the other party. The active party then computes the fully encrypted union and derives the mapping between each local record and the corresponding universal index. This phase finalizes the secure exchange while maintaining the commutativity property, which prevents either participant from learning the intersection membership.

Let us also define $\hat{e}_s$ as the operation of applying $e_s$ to an entire set of identifiers (for example, an entire party’s set $I_k$) and also randomly permuting the order of identifiers. Specifically, for every $k=0,\dots,P-1$,

\begin{equation*}
\widehat{e}_{s}:\prod_{k=0}^{N_k-1}\left\{\prod_{r=1}^{d_{\mbox{\tiny{match}}}}\left[\prod_{l=1}^{L_r-n_r+1}\mathbb{Z}_p\right]\right\}\longrightarrow \prod_{k=0}^{N_k-1}\left\{\prod_{r=1}^{d_{\mbox{\tiny{match}}}}\left[\prod_{l=1}^{L_r-n_r+1}\mathbb{Z}_p\right]\right\}
\end{equation*}
\begin{equation*}
\left(x_{i,r,l}\right)_{i \in \left\{0,\dots,N_k-1\right\}, r\in \left\{1,\dots,d_{\mbox{\tiny{match}}}\right\}, l\in \left\{1,\dots,L_r-n_r+1\right\}}\longmapsto \left(x_{\sigma\left(i\right),r,l}^s\right)_{i \in \left\{0,\dots,N_k-1\right\}, r\in \left\{1,\dots,d_{\mbox{\tiny{match}}}\right\}, l\in \left\{1,\dots,L_r-n_r+1\right\}},
\end{equation*}
where $\sigma:\left\{0,\dots,N_k-1\right\}\longrightarrow \left\{0,\dots,N_k-1\right\}$ is a permutation.

We allow each party to re-index (shuffle) its set after encryption to hide any information that could be inferred from the ordering of identifiers (for example, if one party’s dataset is sorted in some way, a shuffle breaks that correlation). We denote by $\sigma$ a random permutation on the index set $\left\{0,\dots,N_k-1\right\}$ used by party $\mathscr{F}_k$ during encryption. The commutativity still holds: applying $\widehat{e}_{s_1}$ and then $\widehat{e}_{s_2}$ (with independent random permutations) to a vector of identifiers yields the same vector as $\widehat{e}_{s_2}$ followed by $\widehat{e}_{s_1}$ – only the overall order might differ, but since we treat the final results as sets, order does affect correctness.

Let us now define the notion of the product of encryption functions. Let $\left\{s_1,\dots,s_n\right\}\subset \left\{0,\dots,q-1\right\}$, for some $n\in \mathbb{N}\setminus \left\{0\right\}$. Let
\begin{equation*}
\prod_{i=1}^{n}e_{s_i}=e_{s_n}\circ \dots\circ e_{s_1}
\end{equation*}
and
\begin{equation*}
\prod_{i=1}^{n}\widehat{e}_{s_i}=\widehat{e}_{s_n}\circ \dots\circ \widehat{e}_{s_1}.
\end{equation*}

Whenever the index range is empty, i.e., when \(n_1>n_2\) in an expression of the form $\prod_{i=n_1}^{n_2} e_{s_i}$, the corresponding composition is defined to be the identity map.

Since the composition of encryption functions is commutative, the product operator $\prod$ defined above is well-posed. Using this commutative encryption scheme, we can construct Algorithm~\ref{alg:psu-order}, which outlines the PSU protocol with order preservation.

\subsection{PSU Protocol with Order (Exact)}
\label{subsec:PrivateSetUnion(PSU), with order}

We now describe the order–preserving variant of our PSU protocol. This corresponds to the exact-hash alignment regime described in the previous Section. In this case, each identifier $id \in I_{\text{raw},k}$ is first tokenized into \(n\)-grams per feature and hashed into \(I_k\subset\prod_{r=1}^{d_{\text{match}}}\prod_{l=1}^{L_r-n_r+1}\!QR(\mathbb Z_p^\ast)\) as in Section~\ref{subsec:Identifiershashing}; the relative order of \(n\)-grams within each feature is treated as semantically meaningful. The goal is to compute a \emph{universal set of encrypted identifiers} \(U\) that represents the union of all parties’ (ordered) hashed identifiers while revealing nothing about intersection membership, together with local mappings \(\varphi_k: I_{\text{raw},k}\!\to\! V\) that align each party’s records to the universal index set \(V\).

Algorithm~\ref{alg:psu-order} depicts the pseudocode of our solution. At a high level, it proceeds in three phases:

\begin{enumerate}[leftmargin=1.6em]
  \item \textbf{Key setup and first commutative pass.} Each party \( \mathscr F_k \) samples exponents \( s_1^k,s_2^k,s_3^k\in\{0,\dots,q{-}1\} \). Using the commutative mapping \(e_{s}(\cdot)= (\cdot)^s \bmod p\) and its setwise/shuffled lift \(\widehat e_s\), the parties cyclically apply \(\widehat e_{s_1^k}\) to every \(I_{k_1}\), yielding \(\big(\prod_{l} \widehat e_{s_1^l}\big) I_{k_1}\) at the initiator \( \mathscr F_{k_1} \) without exposing which tokens are shared.
  \item \textbf{Provisional union and re-randomization.} The active party \( \mathscr F_{P-1} \) forms the duplicate-free provisional union \( I_{\text{un,prov}} = \bigcup_{k} \big(\prod_l \widehat e_{s_1^l}\big) I_k \) and encodes it by applying \(\prod_l \widehat e_{s_3^l s_2^l}\), producing the final \emph{encrypted universal identifiers} \( U = \big(\prod_l \widehat e_{s_3^l s_2^l}\big) I_{\text{un,prov}} \), which are then broadcast to all parties.
  \item \textbf{Private matching and index assignment.} Each \( \mathscr F_k \) locally re-encrypts its own \(I_k\) with \(e_{s_2^k}\) and relays the result through a second commutative pass of exponents \(\{s_1^\ell,s_2^\ell,s_3^\ell\}\). By commutativity, every element of \(I_k\) is transformed into the same ciphertext as its counterpart in \(U\), enabling \( \mathscr F_k \) to determine \(\varphi_k\) by testing membership in \(U\) \emph{without} revealing whether any particular identifier belongs to the intersection.
\end{enumerate}

The ordered treatment of tokens ensures that two identifiers match if and only if all their feature-wise \(n\)-gram sequences coincide, yielding exact row alignment when inputs are consistently formatted. Security follows from applying only exponentiations in the subgroup \(QR(\mathbb Z_p^\ast)\) under the decisional Diffie–Hellman assumption, together with full-set shuffling \(\widehat e_s\) and a final joint re-randomization, which hides intersection membership while preserving the union. The dominant cost arises from modular exponentiations during the two commutative passes; as noted below, these operations are parallel and can be executed efficiently in a multi-threaded or distributed manner.


\begin{algorithm}[H]
\small
	\SetKwData{Left}{left}\SetKwData{This}{this}\SetKwData{Up}{up}
	\SetKwFunction{Union}{Union}\SetKwFunction{FindCompress}{FindCompress}
	\KwData{For every $k=0,\dots,P-1$, a set of hashed identifiers $I_{k}\subset \prod_{r=1}^{d_{\mbox{\tiny{match}}}}\left[\prod_{l=1}^{L_r-n_r+1}QR\left(\mathbb{Z}_p^{\star}\right)\right]$, with $p$ prime and $q\coloneqq \frac{p-1}{2}$ prime. All the operations on identifiers will be taken modulo $p$. All the operation on indices $k\in \left\{0,\dots,P-1\right\}$ will be taken modulo $P$.}
	\KwResult{Set $V$ of universal indices (UI) and, for any $k=0,\dots,P-1$, map
		\begin{equation*}
		\varphi_k:I_{\mbox{\tiny{raw}},k}\longrightarrow V
		\end{equation*}
		associating each identifier $\mbox{id}\in I_{\mbox{\tiny{raw}},k}$ (of the local dataset $\mathcal{D}_{k}$) to the corresponding index in the set of universal indices (UI) $V$.}
	\BlankLine
	\emph{Initialization}.\\
	\For{$k\leftarrow 0$ \KwTo $P-1$}{
		party $\mathscr{F}_k$ randomly generates three secret exponents $s_1^k$, $s_2^k$ and $s_3^k$ in $\left\{0,\dots,q-1\right\}$. These will be used in different rounds of encryption.
	}
	\emph{First Round – Commutative Encryption of Local Sets}.\\
	\For{$k_1\leftarrow 0$ \KwTo $P-1$}{
		party $\mathscr{F}_{k_1}$ computes
		\begin{equation*}
		\widehat{e}_{s_1^{k_1}}I_{k_1},
		\end{equation*}
		and sends it to the next party $\mathscr{F}_{k_1+1}$ (indices mod $P$ so that e.g. $\mathcal{F}_P$ is $\mathcal{F}_0$).\\
		We now start an additional encryption pipeline, where each subsequent party $\mathcal{F}_{k_2}$ (for $k_2=k_1+1$ up to $P-1$ and then wrapping around to $0,\dots,k_1-1$) encrypts as follows.\\
		\For{$k_2\leftarrow 0,k_2,\neq k_1$ \KwTo $P-1$}{
			When party $\mathscr{F}_{k_2}$ receives $\left[\prod_{l=0,l\neq k_1}^{k_2-1} \widehat{e}_{s_1^{l}}\right]\widehat{e}_{s_1^{k_1}}I_{k_1}$, it computes
			\begin{equation*}
			\left[\prod_{l=0,l\neq k_1}^{k_2} \widehat{e}_{s_1^{l}}\right]\widehat{e}_{s_1^{k_1}}I_{k_1},
			\end{equation*}
			and sends it to $\mathscr{F}_{k_2+1}$.\\
			When $\mathscr{F}_{k_1}$ receives back $\prod_{l=0}^{P-1} \widehat{e}_{s_1^{l}}I_{k_1}$, it stops and sends it to $\mathscr{F}_{P-1}$.
		}
	}
	\emph{Union Computation (Provisional)}.\\
	$\mathscr{F}_{P-1}$ performs the union (repetitions must be avoided)
	\begin{equation*}
	I_{\mbox{\tiny{un,prov}}} \coloneqq \bigcup_{k=0}^{P-1}\prod_{l=0}^{P-1} \widehat{e}_{s_1^{l}}I_{k}.
	\end{equation*}
	Now the active party $\mathcal{F}_{P-1}$ orchestrates a second round of encryption on the provisional union to further blind the values.\\
	\For{$k\leftarrow P-1$ \KwTo $0$}{
		party $\mathscr{F}_{k}$ computes
		\begin{equation*}
		\prod_{l=k}^{P-1} \widehat{e}_{s_3^{l} s_2^{l}}I_{\mbox{\tiny{un,prov}}},
		\end{equation*}
		and sends it to $\mathscr{F}_{k-1}$.
	}
	Once $\mathscr{F}_{P-1}$ receives
	\begin{equation*}
	\prod_{l=0}^{P-1} \widehat{e}_{s_3^{l} s_2^{l}}I_{\mbox{\tiny{un,prov}}},
	\end{equation*}
	party $\mathscr{F}_{P-1}$ determine $U$, the set of encrypted universal identifiers (UID)
	\begin{equation*}
	U\coloneqq \prod_{l=0}^{P-1} \widehat{e}_{s_3^{l} s_2^{l}}I_{\mbox{\tiny{un,prov}}}.
	\end{equation*}
	\caption{PSU protocol, with order.}
    \label{alg:psu-order}
\end{algorithm}

\begin{algorithm}[H]
\small
	\SetKwData{Left}{left}\SetKwData{This}{this}\SetKwData{Up}{up}
	\SetKwFunction{Union}{Union}\SetKwFunction{FindCompress}{FindCompress}
	\BlankLine
	\LinesNumbered
	\setcounter{AlgoLine}{15}
	\For{$k\leftarrow 0$ \KwTo $P-1$}{
		Party $\mathscr{F}_{P-1}$ sends the set of encrypted universal identifiers (UID) to party $\mathscr{F}_{k}$.
	}
	\emph{Private Matching of Identifiers}.\\
	Finally, each party determines the mapping $\phi_k$ between its local identifiers and the universal set. This is done by each party independently (in parallel) using the secret exponents and the encrypted data.\\
	\For{$k_1\leftarrow 0$ \KwTo $P-1$}{
		for any $x\in I_{k_1}$, party $\mathscr{F}_{k_1}$ computes $e_{s_2^{k_1}}x$ and sends it to $\mathscr{F}_{k_1+1}$.\\
		\For{$k\leftarrow 1$ \KwTo $P-1$}{
			When $\mathscr{F}_{k_1+k}$ receives
			\begin{equation*}
			e_{s_2^{k_1}}\prod_{l=1}^{k-1}\left(e_{s_3^{k_1+l}s_2^{k_1+l}s_1^{k_1+l}}\right)x,
			\end{equation*}
			it computes
			\begin{equation*}
			e_{s_2^{k_1}}\prod_{l=1}^{k}\left(e_{s_3^{k_1+l}s_2^{k_1+l}s_1^{k_1+l}}\right)x,
			\end{equation*}
			and sends it to $\mathscr{F}_{k_1+k}$.
			When $\mathscr{F}_{k_1}$ receives back
			\begin{equation*}
			e_{s_2^{k_1}}\prod_{k_2=0, \ k_2\neq k_1}^{P-1}\left(e_{s_3^{k_2}s_2^{k_2}s_1^{k_2}}\right)x,
			\end{equation*}
			party $\mathscr{F}_{k_1}$ computes
			\begin{equation*}
			\prod_{k_2=0}^{P-1}\left(e_{s_3^{k_2}s_2^{k_2}s_1^{k_2}}\right)x
			\end{equation*}
			and party $\mathscr{F}_{k_1}$ stores
			\begin{equation*}
			\tilde{\varphi}_{k_1}:I_{k_1}\longrightarrow U
			\end{equation*}
			\begin{equation*}
			x\longmapsto \prod_{k_2=0}^{P-1}\left(e_{s_3^{k_2}s_2^{k_2}s_1^{k_2}}\right)x
			\end{equation*}
			and determines
			\begin{equation*}
			\varphi_{k_1}:I_{k_1}\longrightarrow V.
			\end{equation*}
		}
	}
		\caption*{PSU protocol, with order (continued).}
\end{algorithm}

At the end of Algorithm~\ref{alg:psu-order}, all parties share a common indexing of the union of identifiers, and none of them has learned which identifiers are exclusively held by which party. The intersection information is protected because any identifier in the intersection appears in $U$ just like any other, with no party knowing if it came from one or multiple datasets.

\textbf{Complexity}. The first round of the algorithm involves each party’s dataset traversing the network of $P$ parties, resulting in $P$ transmissions per dataset (total transmissions of $P^2$ in the worst case). The second round similarly involves $P$ transmissions of the union (which size $N$ could be larger than individual set sizes). The final matching step involves each identifier being sent through $P$ parties (so $N_k \times P$ operations for party $k$). The computational cost is dominated by the modular exponentiations on possibly large sets; however, these are commutative (no interactive OT or public-key operations per item beyond exponentiation).
Since the modular exponentiations on each token are independent, the protocol can be efficiently parallelized across threads or distributed parties. 


In large-scale deployments, a lightweight pre-alignment or ``blocking'' stage could be introduced before the cryptographic protocol to reduce the number of candidate comparisons. For example, each party could exchange salted hash summaries (e.g., SHA-256) of identifiers to exclude obviously non-matching entries, thereby decreasing computational load while preserving privacy. 

\subsection{PSU Protocol without Order (Noisy)}
\label{subsec:PrivateSetUnion(PSU), without order}

Next, we describe the variant of the protocol that does not preserve the order of $n$-grams within each identifier’s features. This corresponds to the more practical ``noisy matching'' alignment regime, designed to handle noisy and heterogeneous identifier fields. This ``without order'' PSU variant is designed to tolerate typographical variations or inconsistencies in identifiers by treating each as an unordered multiset of tokens. In this case, two identifiers can be considered a match (representing the same entity) even if their tokens are in a different order or one identifier has an extra token that the other lacks, as long as a majority of tokens overlap. This is essentially a private noisy matching of identifiers.

The overall structure of the protocol remains similar to the with-order case, but there are two key differences.
\begin{itemize}
	\item In the encryption steps, we introduce an additional random permutation of token positions within each identifier’s feature. Previously, $e_s$ mapped each token but kept its position fixed in its feature. Now, we modify $e_s$ to also randomly permute the positions of the $n$-grams in each feature (or use a fixed permutation $\nu_r$ per exponent) so that the token order information is eliminated (see Definition \eqref{def_encryption_oneexample_without order}).
	
	\item In the union matching step, because tokens are now unordered, we cannot simply take identical encrypted identifiers as an one-to-one match. Two identifiers that represent the same entity might not encrypt to an identical tuple if one had an extra token or tokens were in a different order originally. To address this, we implement a special comparison sub-protocol (Algorithm \ref{compare}) which privately tests if two encrypted identifiers approximately match, given a tolerance threshold $\lambda$. This comparison algorithm essentially counts the number of encrypted tokens two identifiers have in common and determines that they are the same if a sufficiently large fraction of tokens match.
\end{itemize}

We introduce a threshold parameter $0 < \lambda \leq 1$ which governs the matching criterion. For each feature $r$, let $L_r - n + 1$ be the total number of $n$-grams for that feature in a fully formatted identifier (assuming no missing tokens). We assume that each identifier has at least $\lceil \lambda (L_r - n + 1)\rceil$ tokens for feature $r$ (this is reasonable if $\lambda$ is, say, $0.8$, ensuring we only consider matches if both have a significant portion of the full token set). Algorithm \ref{compare} takes two encrypted identifiers (each a set of tokens per feature) and returns $1$ if they are deemed a match (same entity) or $0$ otherwise, without revealing any additional information.

Let $s\in \left\{0,\dots,q-1\right\}$ and
\begin{equation}\label{def_encryption_oneexample_without order}
e_{s}:\prod_{r=1}^{d_{\mbox{\tiny{match}}}}\left[\prod_{l=1}^{L_r-n_r+1}\mathbb{Z}_p\right]\longrightarrow \prod_{r=1}^{d_{\mbox{\tiny{match}}}}\left[\prod_{l=1}^{L_r-n_r+1}\mathbb{Z}_p\right]
\end{equation}
\begin{equation*}
\left(x_{r,l}\right)_{r\in \left\{1,\dots,d_{\mbox{\tiny{match}}}\right\}, l\in \left\{1,\dots,L_r-n_r+1\right\}}\longmapsto  \left(x_{r,\nu_{r}\left(l\right)}^s\right)_{r\in \left\{1,\dots,d_{\mbox{\tiny{match}}}\right\}, l\in \left\{1,\dots,L_r-n_r+1\right\}}
\end{equation*}
and, for every $k=0,\dots,P-1$,
\begin{equation*}
\widehat{e}_{s}:\prod_{j=0}^{N_k-1}\left\{\prod_{r=1}^{d_{\mbox{\tiny{match}}}}\left[\prod_{l=1}^{L_r-n_r+1}\mathbb{Z}_p\right]\right\}\longrightarrow \prod_{j=0}^{N_k-1}\left\{\prod_{r=1}^{d_{\mbox{\tiny{match}}}}\left[\prod_{l=1}^{L_r-n_r+1}\mathbb{Z}_p\right]\right\}
\end{equation*}
\begin{equation*}
\left(x_{j,r,l}\right)_{j \in \left\{0,\dots,N_k-1\right\}, r\in \left\{1,\dots,d_{\mbox{\tiny{match}}}\right\}, l\in \left\{1,\dots,L_r-n_r+1\right\}}\longmapsto \\ \left(x_{\sigma\left(j\right),r,\nu_{r}\left(l\right)}^s\right)_{j \in \left\{0,\dots,N_k-1\right\}, r\in \left\{1,\dots,d_{\mbox{\tiny{match}}}\right\}, l\in \left\{1,\dots,L_r-n_r+1\right\}}
\end{equation*}
with $\nu_{r}:\left\{1,\dots,L_r-n_r+1\right\}\longrightarrow \left\{1,\dots,L_r-n_r+1\right\}$ and $\sigma:\left\{0,\dots,N_k-1\right\}\longrightarrow \left\{0,\dots,N_k-1\right\}$ random permutations (the index $r\in \left\{1,\dots,d_{\mbox{\tiny{match}}}\right\}$).

Here $\nu_r$ is a permutation of the positions for feature $r$. This essentially jumbles the token order for each feature before applying the exponent. Each party can choose a random $\nu_r$ when applying its exponent, or a deterministic one, such as sorting by token value; the important part is that the order is not preserved through encryption.

By the commutativity of the product in $\mathbb{Z}_p$, $e_{s_1}\circ e_{s_2}\simeq e_{s_2}\circ e_{s_1}$, for some $s_1$, $s_2$ in $\left\{0,\dots,q-1\right\}$, the symbol $\simeq$ meaning equality up to a permutation of the $n$-grams. Namely, for $r\in \left\{1,\dots,d_{\mbox{\tiny{match}}}\right\}$, there exist a permutation $\omega_{r}:\left\{1,\dots,L_r-n_r+1\right\}\longrightarrow \left\{1,\dots,L_r-n_r+1\right\}$, such that, for any
\begin{equation*}
\left(x_{r,l}\right)_{r\in \left\{1,\dots,d_{\mbox{\tiny{match}}}\right\}, l\in \left\{1,\dots,L_r-n_r+1\right\}}\in \prod_{r=1}^{d_{\mbox{\tiny{match}}}}\left[\prod_{l=1}^{L_r-n_r+1}\mathbb{Z}_p\right],
\end{equation*}
we have
\begin{equation}\label{def_equivalence}
	e_{s_1}\circ e_{s_2}\left(\left(x_{r,l}\right)_{r\in \left\{1,\dots,d_{\mbox{\tiny{match}}}\right\}, l\in \left\{1,\dots,L_r-n_r+1\right\}}\right)= \\ e_{s_2}\circ e_{s_1}\left(\left(x_{r,\omega_{r}\left(l\right)}\right)_{r\in \left\{1,\dots,d_{\mbox{\tiny{match}}}\right\}, l\in \left\{1,\dots,L_r-n_r+1\right\}}\right)
\end{equation}

Let $\left\{s_1,\dots,s_n\right\}\subset \left\{0,\dots,q-1\right\}$, for some $n\in \mathbb{N}\setminus \left\{0\right\}$. Define
\begin{equation*}
\prod_{i=1}^{n}e_{s_i}=e_{s_n}\circ \dots\circ e_{s_1}
\end{equation*}
and
\begin{equation*}
\prod_{i=1}^{n}\widehat{e}_{s_i}=\widehat{e}_{s_n}\circ \dots\circ \widehat{e}_{s_1}.
\end{equation*}
In products $\prod_{i=n_1}^{n_2}\dots$, if $n_1> n_2$, the result is defined as the identity.

Let $\widehat{N}\coloneqq \sum_{k=0}^{n+1}N_{k}$. We define the concatenation operator as
\begin{equation*}
\Lambda:\prod_{k=0}^{P-1}\left\{\prod_{j=0}^{N_k-1}\left[\prod_{r=1}^{d_{\mbox{\tiny{match}}}}\left(\prod_{l=1}^{L_r-n_r+1}\mathbb{Z}_p\right)\right]\right\}\longrightarrow \prod_{j=0}^{\widehat{N}-1}\left[\prod_{r=1}^{d_{\mbox{\tiny{match}}}}\left(\prod_{l=1}^{L_r-n_r+1}\mathbb{Z}_p\right)\right]
\end{equation*}
\begin{figure*}
\begin{equation*}
\left(\left(x_{i,j,r,l}\right)_{i \in \left\{0,\dots,n\right\},j \in \left\{0,\dots,N_k-1\right\}, r\in \left\{1,\dots,d_{\mbox{\tiny{match}}}\right\}, l\in \left\{1,\dots,L_r-n_r+1\right\}}\right)\longmapsto \left(\tilde{x}_{j,r,l}\right)_{j \in \left\{0,\dots,\widehat{N}-1\right\}, r\in \left\{1,\dots,d_{\mbox{\tiny{match}}}\right\}, l\in \left\{1,\dots,L_r-n_r+1\right\}},
\end{equation*}
\end{figure*}
and $\tilde{x}_{j,r,l}=x_{i,\widehat{j},r,l}$, by setting $\widehat{j}\coloneqq j-\sum_{k=0}^{k-1}N_{k}$ and for
\begin{gather*}
k\in \left\{0,\dots,P-1\right\}, \\
j \in \left\{\sum_{k=0}^{k-1}N_{k},\dots,\sum_{k=0}^{i}N_{k}-1\right\}, \\ 
r\in \left\{1,\dots,d_{\mbox{\tiny{match}}}\right\}, \\
l\in \left\{1,\dots,L_r-n_r+1\right\}.
\end{gather*}

Before defining the main Algorithm \ref{Alg_PSU_1_without_order}, we need to define an algorithm for comparing two encrypted identifiers as shown below.

\begin{algorithm}[H]
\small
	\SetKwData{Left}{left}\SetKwData{This}{this}\SetKwData{Up}{up}
	\SetKwFunction{Union}{Union}\SetKwFunction{FindCompress}{FindCompress}
	\KwData{Encrypted identifiers
		\begin{equation*}
		\left(x_{r,l}\right)_{r\in \left\{1,\dots,d_{\mbox{\tiny{match}}}\right\}, l\in  \left\{1,\dots,p_{1,r}\right\}}\in \prod_{r=1}^{d_{\mbox{\tiny{match}}}}\left[\prod_{l=1}^{p_{1,r}}\mathbb{Z}_p\right]
		\end{equation*}
		and
		\begin{equation*}
		\left(y_{r,l}\right)_{r\in \left\{1,\dots,d_{\mbox{\tiny{match}}}\right\}, l\in \left\{1,\dots,p_{2,r}\right\}}\in \prod_{r=1}^{d_{\mbox{\tiny{match}}}}\left[\prod_{l=1}^{p_{2,r}}\mathbb{Z}_p\right].
		\end{equation*}
		Threshold $\lambda \in \left(0,1\right]$. Assume $p_{i,r}\in \left[ \lceil \lambda L_r-n_r+1\rceil, L_r-n_r+1\right]$, for $r\in \left\{1,\dots,d_{\mbox{\tiny{match}}}\right\}$.}
	\KwResult{$1$ if the encrypted identifiers match. $0$ otherwise. If $1$, return $\mbox{list}_{\mbox{\tiny{match}}}$.}
	\BlankLine
	\emph{Initialization}.\\
	Set
	\begin{equation*}
	\mbox{list}_{\mbox{\tiny{match}}} \leftarrow \left[\, \right].
	\end{equation*}
	\emph{Comparison}.\\
	\For{$r\leftarrow 1$ \KwTo $d_{\mbox{\tiny{match}}}$}{
		Set
		\begin{equation*}
		\mbox{list}_{\mbox{\tiny{match}},r} \leftarrow \left[\, \right].
		\end{equation*}
			\For{$l_1\leftarrow 0$ \KwTo $p_{1,r}$}{
				\For{$l_2\leftarrow 0$ \KwTo $p_{2,r}$}{
					If
					\begin{equation*}
						x_{r,l_1} = y_{r,l_2}
					\end{equation*}
					and $l_1$ does not belong to $\mbox{list}_{\mbox{\tiny{match}},r}$, append $l_1$ to $\mbox{list}_{\mbox{\tiny{match}},r}$.
				}
			}
			If the cardinality of $\mbox{list}_{\mbox{\tiny{match}},r}$ is greater than
			\begin{equation*}
				\lceil \lambda L_r-n_r+1\rceil,
			\end{equation*}
			return $0$.
			Else, append $\mbox{list}_{\mbox{\tiny{match}},r}$ to $\mbox{list}_{\mbox{\tiny{match}}}$.
		}
	Return $1$ and $\mbox{list}_{\mbox{\tiny{match}}}$.
	\caption{Encrypted Identifier Comparison.}\label{compare}
\end{algorithm}

\begin{remark}[Equivalence]\label{def_rel}
	Observe that algorithm \eqref{compare} defines a relation between
	\begin{equation*}
		\prod_{r=1}^{d_{\mbox{\tiny{match}}}}\left[\prod_{l=1}^{p_{1,r}}\mathbb{Z}_p\right]
	\end{equation*}
	and
	\begin{equation*}
		\prod_{r=1}^{d_{\mbox{\tiny{match}}}}\left[\prod_{l=1}^{p_{2,r}}\mathbb{Z}_p\right],
	\end{equation*}
	where two respective elements
	\begin{equation*}
	\*x\coloneqq \left(x_{r,l}\right)_{r\in \left\{1,\dots,d_{\mbox{\tiny{match}}}\right\}, l\in \left\{1,\dots,p_{1,r}\right\}}
	\end{equation*}
	and
	\begin{equation*}
	\*y\coloneqq \left(y_{r,l}\right)_{r\in \left\{1,\dots,d_{\mbox{\tiny{match}}}\right\}, l\in \left\{1,\dots,p_{2,r}\right\}}
	\end{equation*}
	are related if algorithm \eqref{compare} applied to $\left(\*x,\*y\right)$ returns $1$. However, this relation may not be an equivalence relation, since it may not be transitive.
\end{remark}

An adaptive selection of the matching threshold $\lambda$ could further improve PPEA accuracy by adjusting it according to data quality or field variability. For instance, domain-specific calibration (e.g., different $\lambda$ values for names and addresses) or automatic tuning using validation data could balance recall and precision in noisy matching without altering the cryptographic design.
Further extensions could integrate alternative similarity metrics, such as edit distance or locality-sensitive hashing (LSH)~\cite{indyk1998approximate} 
, enabling matching beyond token overlap while maintaining privacy-preserving properties.

Next, we present the main algorithm for our PSU without order.

\begin{algorithm}[H]
\small
	\SetKwData{Left}{left}\SetKwData{This}{this}\SetKwData{Up}{up}
	\SetKwFunction{Union}{Union}\SetKwFunction{FindCompress}{FindCompress}
	\KwData{For every $k=0,\dots,P-1$, a set of hashed identifiers $I_{k}\subset \prod_{r=1}^{d_{\mbox{\tiny{match}}}}\left[\prod_{l=1}^{L_r-n_r+1}QR\left(\mathbb{Z}_p^{\star}\right)\right]$, with $p$ prime and $q\coloneqq \frac{p-1}{2}$ prime. All the operations on identifiers will be taken modulo $p$. All the operation on indices $k\in \left\{0,\dots,P-1\right\}$ will be taken modulo $P$. Threshold $\lambda \in \left(0,1\right]$.}
	\KwResult{Set $V$ of universal indices (UI) and, for any $k=0,\dots,P-1$, map
		\begin{equation*}
		\varphi_k:\left\{0,\dots,N_k-1\right\}\longrightarrow V
		\end{equation*}
		associating each index $i$ of the local dataset $\mathcal{D}_{k}$ to the corresponding index in the set of universal indices (UI) $V$.}
	\BlankLine
	\emph{Initialization}.\\
	\For{$k\leftarrow 0$ \KwTo $P-1$}{
		party $\mathscr{F}_k$ randomly generates $s_1^k$, $s_2^k$ and $s_3^k$ in $\left\{0,\dots,q-1\right\}$.
	}
	\emph{First Round – Commutative Encryption of Local Sets}.\\
	\For{$k_1\leftarrow 0$ \KwTo $P-1$}{
		party $\mathscr{F}_{k_1}$ computes
		\begin{equation*}
		\widehat{e}_{s_1^{k_1}}I_{k_1},
		\end{equation*}
		and sends it to $\mathscr{F}_{k_1+1}$.\\
		\For{$k_2\leftarrow 0,k_2,\neq k_1$ \KwTo $P-1$}{
			When $\mathscr{F}_{k_2}$ receives $\left[\prod_{l=0,l\neq k_1}^{k_2-1} \widehat{e}_{s_1^{l}}\right]\widehat{e}_{s_1^{k_1}}I_{k_1}$, it computes
			\begin{equation*}
			\left[\prod_{l=0,l\neq k_1}^{k_2} \widehat{e}_{s_1^{l}}\right]\widehat{e}_{s_1^{k_1}}I_{k_1},
			\end{equation*}
			and sends it to $\mathscr{F}_{k_2+1}$.\\
			When $\mathscr{F}_{k_1}$ receives back $\prod_{l=0}^{P-1} \widehat{e}_{s_1^{l}}I_{k_1}$, it stops and sends it to $\mathscr{F}_{P-1}$.
		}
	}
	\emph{Union Computation (Provisional)}.\\
	$\mathscr{F}_{P-1}$ define a vector 
	\begin{equation*}
	V_{\mbox{\tiny{un,prov}}} \coloneqq \Lambda\left(\prod_{l=0}^{P-1} \widehat{e}_{s_1^{l}}I_{0},\dots,\prod_{l=0}^{P-1} \widehat{e}_{s_1^{l}}I_{k},\dots,\prod_{l=0}^{P-1} \widehat{e}_{s_1^{l}}I_{n}\right).
	\end{equation*}
	We have
	\begin{equation}\label{hatN_linked_length}
		\widehat{N}= \mbox{length}_{\mbox{first axis}}\left(V_{\mbox{\tiny{un,prov}}}\right).
	\end{equation}
	\For{$k_1\leftarrow 0$ \KwTo $\widehat{N}-1$}{
		\For{$k_2\leftarrow k_1$ \KwTo $\widehat{N}-1$}{
		$\mathscr{F}_{P-1}$ applies algorithm \ref{compare} to $V_{\mbox{\tiny{un,prov}},k_1}$, with $V_{\mbox{\tiny{un,prov}},k_2}$. If the output is $1$, remove $V_{\mbox{\tiny{un,prov}},k_2}$ and remove from $V_{\mbox{\tiny{un,prov}},k_1}$ $n$-grams indicated in $\mbox{list}_{\mbox{\tiny{match}}}$.
		}
	}
    From elements of $V_{\mbox{\tiny{un,prov}},k_1}$, party $\mathscr{F}_{P-1}$ removes randomly components to reduce length of each element to
	$\lceil \lambda L_r-n_r+1\rceil$. $\mathscr{F}_{P-1}$ defines
	\begin{equation*}
		I_{\mbox{\tiny{un,prov}}} \coloneqq \left\{V_{\mbox{\tiny{un,prov}},l} \ \Big| \ l\in \left\{1,\dots,\text{length}_{\text{\scriptsize first axis}}\left(V_{\text{\tiny{un,prov}}}\right)\right\}\right\},
	\end{equation*}
	where $\text{length}_{\text{\scriptsize first axis}}\left(V_{\text{\scriptsize{un,prov}}}\right)$ may have changed, with respect to \eqref{hatN_linked_length}, because of the removals.\\
    \For{$k\leftarrow P-1$ \KwTo $0$}{
		party $\mathscr{F}_{k}$ computes
		\begin{equation*}
		\prod_{l=k}^{P-1} \widehat{e}_{s_3^{l} s_2^{l}}I_{\mbox{\tiny{un,prov}}},
		\end{equation*}
		and sends it to $\mathscr{F}_{k-1}$.
	}
    Once $\mathscr{F}_{P-1}$ receives
	\begin{equation*}
	\prod_{l=0}^{P-1} \widehat{e}_{s_3^{l} s_2^{l}}I_{\mbox{\tiny{un,prov}}},
	\end{equation*}
	\caption{PSU protocol, without order.}\label{Alg_PSU_1_without_order}
\end{algorithm}

\begin{algorithm}
	\SetKwData{Left}{left}\SetKwData{This}{this}\SetKwData{Up}{up}
	\SetKwFunction{Union}{Union}\SetKwFunction{FindCompress}{FindCompress}
	\BlankLine
	\LinesNumbered
	\setcounter{AlgoLine}{18}
	
	\For{$i\leftarrow 1$ \KwTo $P-1$}{
		party $\mathscr{F}_{P-1}$ sends
		\begin{equation*}
		\prod_{l=0}^{P-1} \widehat{e}_{s_3^{l} s_2^{l}}I_{\mbox{\tiny{un,prov}}},
		\end{equation*}
		to $\mathscr{F}_{k}$.
	}
	Party $\mathscr{F}_{P-1}$ determines $U$ the set of encrypted universal identifiers (UID)
	\begin{equation*}
	U\coloneqq \prod_{l=0}^{P-1} \widehat{e}_{s_3^{l} s_2^{l}}I_{\mbox{\tiny{un,prov}}}.
	\end{equation*}
	\For{$k\leftarrow 0$ \KwTo $P-1$}{
		Party $\mathscr{F}_{P-1}$ sends the set of encrypted universal identifiers (UID) to party $\mathscr{F}_{k}$.
	}
	\emph{Private Matching of Identifiers}.\\
	\For{$k_1\leftarrow 0$ \KwTo $P-1$}{
		for any $x\in I_{k_1}$, party $\mathscr{F}_{k_1}$ computes $e_{s_2^{k_1}}x$ and sends it to $\mathscr{F}_{k_1+1}$.\\
		\For{$k\leftarrow 1$ \KwTo $P-1$}{
			When $\mathscr{F}_{k_1+k}$ receives
			\begin{equation*}
			e_{s_2^{k_1}}\prod_{l=1}^{k-1}\left(e_{s_3^{k_1+l}s_2^{k_1+l}s_1^{k_1+l}}\right)x,
			\end{equation*}
			it computes
			\begin{equation*}
			e_{s_2^{k_1}}\prod_{l=1}^{k}\left(e_{s_3^{k_1+l}s_2^{k_1+l}s_1^{k_1+l}}\right)x,
			\end{equation*}
			and sends it to $\mathscr{F}_{k_1+k}$.
			When $\mathscr{F}_{k_1}$ receives back
			\begin{equation*}
			e_{s_2^{k_1}}\prod_{k_2=0, \ k_2\neq k_1}^{P-1}\left(e_{s_3^{k_2}s_2^{k_2}s_1^{k_2}}\right)x,
			\end{equation*}
			party $\mathscr{F}_{k_1}$ computes
			\begin{equation}\label{encr_id}
			\prod_{k_2=0}^{P-1}\left(e_{s_3^{k_2}s_2^{k_2}s_1^{k_2}}\right)x.
			\end{equation}
			Party $\mathscr{F}_{k_1}$ checks the existence of an element $\hat{x}\in U$, having $\lceil \lambda L_r-n_r+1\rceil$ $n$-grams in common, with \eqref{encr_id}. If this element exists, party $\mathscr{F}_{k_1}$ stores
			\begin{equation*}
			\tilde{\varphi}_{k_1}:I_{k_1}\longrightarrow U
			\end{equation*}
			\begin{equation*}
			x\longmapsto \hat{x}
			\end{equation*}
			and determines
			\begin{equation*}
			\varphi_{i_1}:I_{i_1}\longrightarrow V.
			\end{equation*}
		}
	}
	\caption*{PSU protocol, without order (continued).}
\end{algorithm}

\begin{remark}\label{remark_implementation}
In case the order of $n$-grams is not preserved, to reduce the computational cost as well as save memory, Bloom filters (bit arrays) are used to store encrypted identifiers, i.e., an encrypted identifier
	\begin{equation}\label{def_repr_bloomfilter}
		\left(x_{r,l}\right)_{r\in \left\{1,\dots,d_{\mbox{\tiny{match}}}\right\}, l\in \left\{1,\dots,L_r-n_r+1\right\}}\in \prod_{r=1}^{d_{\mbox{\tiny{match}}}}\left[\prod_{l=1}^{L_r-n_r+1}\mathbb{Z}_p\right],
	\end{equation}
	is represented by a Bloom filter (bitarray) $\*b\in \left\{0,1\right\}^{p}$ by setting
	\begin{equation*}
		\*b\left(x_{r,l}\right) = 1, \forall \left(r,l\right)\in \left\{1,\dots,d_{\mbox{\tiny{match}}}\right\}\times \left\{1,\dots,L_r-n_r+1\right\}
	\end{equation*}
	and $0$ elsewhere. We highlight two properties of this representation:
	\begin{enumerate}
		\item Two identifiers that are equivalent according to \eqref{def_equivalence} produce the same Bloom filter, since Bloom filters defined in \eqref{def_repr_bloomfilter} are insensitive to $n$-gram order.
		\item Two identifiers that are not equivalent according to \eqref{def_equivalence} may still produce the same Bloom filter, because the Bloom filters in \eqref{def_repr_bloomfilter} ignore both $n$-gram order and feature order. This may lead to false matches. Using a separate Bloom filter per feature would reduce this effect, but here we use a single Bloom filter for all features to save memory.
	\end{enumerate}
\end{remark}

In future implementations, precision could be further improved by adopting multiple Bloom filters (one per identifier feature) or by applying multiple hash functions per token, thereby reducing false positives while maintaining reasonable memory use.

In scenarios where a subset of entities already share persistent global identifiers (e.g., national or organizational IDs), these records can be excluded from the PSU protocol and directly merged, with the alignment executed only on the remaining unmatched entities. This optimization preserves security while avoiding redundant computation in practical deployments.
An additional layer of formal privacy could be incorporated by introducing DP perturbations, such as dummy identifiers or randomized mappings, to protect against membership inference in extreme asymmetric cases.

Finally, Bloom filter construction can be parallelized over data blocks, offering substantial runtime reductions for large-scale datasets, as shown in other parallel PSU implementations such as PULSE~\cite{gao2025pulse}.

\subsection{Synthetic Data Completion}

During model training, each party must generate a synthetic dataset to fill in the missing features in its local data, i.e., those present in the datasets of other parties but absent locally. To this end, the Synthetic Data Vault (SDV)~\cite{7796926} is the state-of-the-art open-source library for generating high-quality synthetic data.

SDV uses ML models to capture the statistical properties and dependencies of real datasets, enabling the creation of synthetic data that preserves both structure and utility, while protecting individual privacy. It supports tabular, time-series, and relational data and provides tools for evaluating data fidelity and privacy. By offering privacy-compliant data, SDV facilitates secure data sharing, model testing, and training in sensitive domains such as healthcare and finance, without compromising confidentiality.
Two SDV backends are particularly suitable for our purpose: Gaussian Copula and Conditional Tabular Generative Adversarial Network (CTGAN).

The Gaussian Copula backend employs statistical modeling through copulas, i.e., functions that describe dependencies between random variables, to generate synthetic data. Each feature is first fitted to an appropriate marginal distribution (e.g., Gaussian, Exponential) and normalized within a uniform range $(0,1)$ via cumulative distribution functions (CDFs). Correlations among variables are captured by a correlation matrix that models the linear dependence between transformed variables. Synthetic samples are then generated from the fitted Gaussian Copula model and inverse-transformed back to the original feature space. By operating in a transformed space and focusing on correlations rather than exact values, this method ensures that individual-level information is not reproduced while maintaining realistic relationships among variables. Gaussian Copula is particularly effective for tabular data of moderate dimensionality, capturing non-linear dependencies with low computational cost.

The CTGAN backend extends the standard Generative Adversarial Network (GAN) architecture to tabular data with mixed types, imbalanced distributions, and/or categorical features. It consists of two neural components: (i) a \emph{generator}, which produces synthetic records, and (ii) a \emph{discriminator}, which distinguishes real from synthetic samples. Continuous features are normalized using min–max scaling and mode-specific normalization, which increases representation in dense data regions, while categorical variables are transformed into binary vectors. CTGAN selects a column at random during training as a conditioning variable and samples data accordingly, allowing the generator to learn relationships between that column and the rest of the dataset. After training, the generator produces realistic synthetic rows from random noise and conditional inputs, which are then inverse-transformed to match the original data types and distributions. This approach performs particularly well when the underlying data relationships are complex and traditional statistical models fail to capture them.

From the perspective of missing-data theory, our synthetic completion step can be viewed as a form of model-based imputation, where unobserved features are generated conditional on the observed ones. Classical approaches distinguish between (i) data imputation, which replaces missing values by point estimates or draws from the posterior predictive distribution (e.g., via the EM algorithm or multiple imputation), and (ii) marginalization, where learning and inference integrate over the distribution of missing values without explicitly filling them in~\cite{dempster1977maximum,rubin1976inference,little2019statistical,schafer1997analysis}. Our PSU-based framework is compatible with both views: the synthetic features we generate correspond to imputations of unobserved modalities, while downstream federated models could in principle be trained in a marginalization style by averaging over multiple synthetic completions, leveraging the rich toolbox developed in the missing-data literature.

\section{Conclusions}
\label{sec:Conclusions}



In this paper, we presented the \sherpa multi-party PSU protocol for PPEA in VFL. The proposed method enables multiple parties to align their datasets without revealing intersection membership, thereby strengthening privacy guarantees in collaborative ML settings. The protocol generalizes the prior two-party PSU approaches to a multi-party scenario with low communication overhead. Two complementary variants were presented: an order-preserving version for exact alignment and an unordered version that supports noisy matching, tolerant to typographical and formatting inconsistencies. Together, these algorithms offer a flexible trade-off between precision and robustness, depending on data quality and application context.

Beyond the core design, we discussed implementation aspects, including Bloom filter representations, adaptive thresholds, and parallelization strategies, to enhance scalability. Through secure and accurate entity alignment, the proposed \sherpa PSU protocol empowers organizations to collaborate on VFL applications in a privacy-preserving manner, enabling practical, privacy-preserving VFL across sensitive domains, including healthcare, finance, manufacturing, aerospace, cybersecurity, and the defense industry.

In practice, both exact and noisy PSUs can be integrated with a paradigm such as SBVFL~\cite{acero2025sherpa}, which establishes a new benchmark in privacy preservation while enhancing computational efficiency. This paradigm achieves stronger confidentiality guarantees than conventional aggregation methods, alongside faster convergence and improved scalability. SBVFL thus represents a safer and more resilient framework for the next generation of secure, privacy-preserving, and collaborative intelligence.



\section*{Contributions and Acknowledgments} \label{app:A}


Daniel M. Jimenez-Gutierrez

Dario Pighin

Enrique Zuazua

Georgios Kellaris

Joaquin Del Rio

Oleksii Sliusarenko

Xabi Uribe-Etxebarria

\vspace{5mm}
The authors are presented in alphabetical order by first name.


\bibliographystyle{abbrv}
\bibliography{my_references_Sherpa.bib}


\end{document}